\documentclass[12pt]{article}

\usepackage{epsf}
\usepackage{amssymb}
\usepackage{rotate}
\newcommand{\ba}{\begin{array}}
\newcommand{\ea}{\end{array}}
\newcommand{\bd}{\begin{displaymath}}
\newcommand{\ed}{\end{displaymath}}
\newcommand{\be}{\begin{equation}}
\newcommand{\ee}{\end{equation}}
\newcommand{\bea}{\begin{eqnarray}}
\newcommand{\eea}{\end{eqnarray}}

\def\a{\alpha}
\def\b{\beta}
\def\g{\gamma}

\def\e{\epsilon}

\def\th13 {\theta_{13}}

\usepackage{graphicx}

\catcode`\@=11
\def\lsim{\mathrel{\mathpalette\@versim<}}
\def\gsim{\mathrel{\mathpalette\@versim>}}
\def\@versim#1#2{\vcenter{\offinterlineskip
\ialign{$\m@th#1\hfil##\hfil$\crcr#2\crcr\sim\crcr } }}
\catcode`\@=12

\catcode`@=12
\begin{document}
\thispagestyle{empty}
\begin{flushright}
\end{flushright}
\vspace{0.3in}
\begin{center}
{\Large \bf Magic baseline and magic energy in neutrino oscillation
with non-standard interactions}\\

\vspace{0.8in}
{\bf Zini Rahman\footnote{E-mail: zini.rahman.s@gmail.com}
and Rathin Adhikari\footnote{E-mail: rathin@ctp-jamia.res.in}\\}
\vspace{0.2in}
{\sl Centre for Theoretical Physics, Jamia Millia Islamia -- Central 
University,\\ Jamia Nagar, 
New Delhi 110025, India\\} 

\end{center}
\vspace{0.8in}
\begin{abstract}\
 We have discussed  conditions under which  probability of oscillation
 ($\nu_e\rightarrow \nu_{\mu}$) is independent
 of CP violating phase $\delta$. The condition
  of magic baseline on its length is well-known. We have  proposed another condition which
  is on
  neutrino energy.
 We have shown that magic baseline condition is not possible in general,
 for small $\theta_{13}$ with non-standard interaction and for large
 $\theta_{13}$
 with both standard and non-standard interactions. However,
 neutrino energy condition is possible for such cases as well as
 for cases where magic baseline condition is applicable. We have discussed
 how one may resolve hierarchy problem for neutrino
  masses by using such energy condition.
 For a baseline of length 650 Km, using this
 energy condition we discuss the possible number of $\mu^-$ events
 at the detector for a period of 5 years
 and
 also the sensitivity in measurement of  $\cos^2 \theta_{13}$.
 \end{abstract}

\newpage
The probability of oscillation of different flavors of neutrinos
depends on various parameters present in
the neutrino mixing matrix - the PMNS matrix \cite{pmns} as well as the mass squared
differences.
Although two angles $\theta_{12}$ and $\theta_{23}$ are known with some accuracy
but there is  only upper bound for $\theta_{13}$ \cite{tho} and CP violating phase
$\delta$ is unknown. Although mass squared differences for different
neutrinos
are known but the sign of ${\Delta m_{31}}^2$ (where $\Delta m_{ij}^2 =
m_i^2 - m_j^2$ and $m_i$ is the mass of $i$-th neutrino)
is unknown.  
Due to correlations among these unknowns there are ambiguities \cite{ki}
in analysing neutrino oscillation data. To reduce such ambiguities it
 is useful to choose suitable baseline \cite{base}.
Particularly, magic baseline \cite{magic1,magic2,magic3,magic4} is
useful for some specific length for which the perturbative expression of  probability
$P_{\nu_e \rightarrow \nu_\mu}$ is independent of $\delta$ upto order $\a^2$
(where $\a = \Delta m^2_{21}/{\Delta m^2_{31}}$).
Then it is easier to find out other parameters
apart from $\delta$ which plays the role in neutrino oscillation.
However, using magic baseline has its  limitations  also
- namely (a) it may not be always
possible to place the detector at a magic baseline distance from the source of
neutrino production, (b) if there is non-standard interaction
(NSI) \cite{nsi0}
then we have shown that it is difficult to get  $\delta$ independent
probability
$P_{\nu_e \rightarrow \nu_\mu}$
using the magic baseline condition.

In this work we present another condition
- which may be termed as magic energy condition under which also the probability
of oscillation will be  independent of $\delta$ upto order
$\a^2$. Considering this condition
one might be able to circumvent the above two shortcomings of the magic baseline.
Unlike magic baseline
condition this condition depends
on length of the baseline also apart from its
dependence on $\sqrt{2} G_F n_e$ (Here $n_e$ is the electron number density
of the matter).
To use the magic energy condition, one option could be to analyse
 oscillation data in small energy bins. However, the other better
 option could be to use monoenergetic neutrino beam at the source as
proposed in recent years \cite{mon, mono}. The idea is about
using nucleus which absorbs an electron and emits a neutrino. By acelerating the mother
nuclei with suitable Lorentz boost
factor one may get the suitable neutrino energy  which is satisfied by
magic energy condition. Due to the monoenergetic nature of the beam it
 is expected to have better precision in finding various neutrino oscillation
 parameters.

 The magic baseline condition was initially obtained using
the
perturbative expansion for small $\theta_{13}$ with Standard Model
 interaction (SMI)
\cite{magic1}. Here, we
obtain the modified form
of magic baseline condition for both small and large $\theta_{13}$ and for
both SMI and NSI.
Besides, we obtain the magic neutrino energy condition for both SMI and NSI
and also  for
both small and large $\theta_{13}$ as allowed by present experiment
\cite{tho}.
Finally, we have compared advantages and disadvantages
in considering magic baseline condition and magic
 energy condition in experiments.

Flavor eigenstates
$\nu_\alpha$ may be related to  mass eigenstates of neutrinos $\nu_i$ as
\be
\vert\nu_\alpha>=\sum_{i}  U_{\alpha i}\vert\nu_i>
, \;\;U=R_{23} R_{13}(\delta) R_{12}
\;\;\quad \textrm{and}
\qquad i=1, 2, 3,
\ee
where $U$ is PMNS matrix \cite{pmns} and $R_{ij}$ are the rotation
matrices.
General probability expression for oscillation of neutrino of flavor $l$ to
neutrino flavor $m$ in matter (satisfying adiabatic condition for the density
of matter) is given by
\bea
 P(\nu_{l}\rightarrow\nu_{m})=\delta_{lm}-4\sum_{i>j}  Re[J_{ij}^{lm}]
\sin^2\Delta^{'}_{ij}+2 \sum_{i>j} Im[J_{ij}^{lm}] \sin 2 \Delta^{'}_{ij}
\eea
where
$ J_{ij}^{lm}=U^{'}_{li} {U^{'\ast}_{lj}} {U^{'\ast}_{mi}} U^{'}_{mj} $ and
$\Delta^{'}_{ij}=
\Delta ^{'}m_{ij}^2 L/ (4E) $. 
Here $ \Delta^{'} m_{ij}^2={m^{'}_{i}}^2-{m^{'}_{j}}^2 $ and label ($\;^{'}\;$) indicates the
neutrino matter interaction induced quantities corresponding to those quantities
in vacuum.
Let us write $x=\Delta^{'}_{31} $ , $y=\Delta^{'}_{32}$
and $z=\Delta^{'}_{12}$. Using trigonometric identities :
$
-\sin^2x +\sin^2y-\sin^2z=2 \sin x\; \cos y\; \sin z
$
 and $
-\sin 2x+\sin 2y-\sin 2z=-4 \sin x\; \sin y\; \sin z
$
where $x,\; y, \; z$ obey the relationship
$ x=y-z , y=x+z $ and $ z=y-x$ 
and putting the condition $ \sin z =0 $,
the probability expression $P_{\nu_e \rightarrow \nu_{\mu}}$
can be written as
\bea
P_{\nu_e \rightarrow \nu_{\mu}}
= -4 \left( Re\left[U^{'}_{13}{U^{'\ast}_{11}} {U^{'\ast}_{23}}
U^{'}_{21}\right]  +
Re\left[U^{'}_{13}{U^{'\ast}_{12}} U^{'\ast}_{23} U^{'}_{22}\right] \right)
\sin^2 \Delta^{'}_{31} \nonumber \\
-2 \left( Im\left[U^{'}_{13}{U^{'\ast}_{11}} {U^{'\ast}_{23}} U^{'}_{21}\right] +
Im\left[U^{'}_{13}{U^{'\ast}_{12}} {U^{'\ast}_{23}} U^{'}_{22}\right] \right) \sin 2
\Delta^{'}_{31}.
\eea
The condition $\sin z = 0 $ implies
\bea
\mid \Delta^{'}_{21} \mid =  \pm n \pi \;
\mbox{where } n \; \mbox{is positive integer.}
\eea
Similarly, under the condition $ \sin x =0 $,
the probability expression $P_{\nu_e \rightarrow \nu_{\mu}}$ 
can be written as 
\bea
P_{\nu_e \rightarrow \nu_{\mu}}
= -4 \left( Re\left[U^{'}_{12}{U^{'\ast}_{11}} {U^{'\ast}_{22}} U^{'}_{21}
\right]  +
Re\left[U^{'}_{13}{U^{'\ast}_{12}} U^{'\ast}_{23} U^{'}_{22}\right] \right)
\sin^2 \Delta^{'}_{12} \nonumber \\ 
-2 \left( Im\left[U^{'}_{12}{U^{'\ast}_{11}} {U^{'\ast}_{22}} U^{'}_{21}
\right] -
Im\left[U^{'}_{13}{U^{'\ast}_{12}} {U^{'\ast}_{23}} U^{'}_{22}\right] \right) \sin
2\Delta^{'}_{12}
\eea
and here the condition $\sin x = 0 $ implies
\bea
\mid \Delta^{'}_{31} \mid = \pm n \pi \; 
\mbox{where } \; n \; \mbox{is positive integer.}
\eea
The condition in eq.(4) or in (6) essentially either fixes the length of
baseline or the energy of the neutrino beam. To determine elements of $U^{'}$
and $\Delta^{'}_{ij}$ we shall use the perturbative approach for small and large
$\theta_{13}$ seperately. 

We discuss in brief the perturbative approach here. The diagonal neutrino mass matrix is approximately given by
\be
m \approx \Delta m_{31}^2 diag(0,\a,1).
\ee
The effective Hamiltonian  induced by interaction of matter with neutrinos
is written in weak interaction basis as
\be
 H \approx \frac{\Delta m_{31}^2}{2E}R_{23}MR_{23}^\dag
\ee
where
\be
 M= R_{13}R_{12}\; {m \over \Delta m_{31}^2}\;R_{12}^\dag R_{13}^\dag
 + diag(A,0,0) + R_{23}^\dag\pmatrix{0 & X & Y \cr X^{\ast} & B & C \cr
 Y^{\ast}
 & C^{\ast} & D} R_{23}\; .
\ee
In equation (9)  
\bea
A=\frac{2E\sqrt{2}G_{F}n_{e}}{\Delta m_{31}^2},\;
 X=\frac{2E\e_{12}}{\Delta m_{31}^2}, \;Y=\frac{2E\e_{13}}{\Delta
 m_{31}^2}, \;
B=\frac{2E\e_{22}}{\Delta m_{31}^2},\;
C=\frac{2E\e_{23}}{\Delta m_{31}^2},\;
D=\frac{2E\e_{33}}{\Delta m_{31}^2},
\eea
where $A$ is considered due to Standard model interaction of neutrinos with
electron. $\e_{12}$ , $\e_{13}$, $\e_{22}$, $\e_{23}$
 and $\e_{33}$  are considered due to NSI
 of neutrinos with matter \cite{nsi0} (e.g, in $R$ violating Supersymmetric
 Models
 neutrinos may interact with down type quarks through squark exchange
 and may interact with electron through slepton exchange \cite{susy} and in
 Minimal Supersymmetric Standard Model with right-handed neutrinos
 \cite{susy1} through lepton number violating interactions accompanied
 with neutrinos). We consider magnitude of $B, C, D, X, Y$ due to NSI
 not higher than $\a$  due to various experimental constraints \cite{coup}. In (9), ($\; ^{*} \;$)  is
 denoted for complex conjugation.
In (10),  $G_{F}$ is the Fermi constant and $n_{e}$ is the electron number density.

The mixing matrix $ U^{\prime}$ can be found out as $ U^{\prime}=
R_{23} \; W $. Here, $W$
is the normalized eigenvectors of $\Delta m_{31}^2M /(2 E)$
calculated through perturbative technique.
We follow the technique adopted  in \cite{perturb} for Standard Model
interactions.
Let us consider the case where only $\e_{12}$ and
$\e_{13}$ are present as NSI
and where $ \sin \theta_{13} $ is 
small and of the order of $\alpha$ or less.
M can be written as $ M = M^{(0)}+M^{(1)}+M^{(2)} $ where $M^{i}$ contains terms
of the order of $\alpha^{i}$.
Then we can write
\bea
 M^{(0)}=\frac{\Delta m_{31}^2}{2E} \; diag(A,0,1), \;\; M^{(1)}=
 \frac{\Delta m_{31}^2}{2E}\pmatrix{\alpha s_{12}^2 & b & a\cr b^{\ast}
 & \alpha c_{12}^2 & 0\cr a^{\ast} & 0 & 0}, \nonumber \\
 M^{(2)}=\frac{\Delta m_{31}^2}{2E}\pmatrix{s_{13}^2 & 0 & -e^{-i\delta}\alpha c_{13} s_{12}^2 s_{13} \cr 0 & 0 & -e^{-i\delta}\alpha c_{12} s_{12} s_{13} \cr -e^{i\delta}\alpha c_{13} s_{12}^2 s_{13} & -e^{i\delta}\alpha c_{12} s_{12} s_{13} & - s_{13}^2}
\eea
where
\bea
  a= c_{23}Y+e^{-i\delta}c_{13}s_{13} +Xs_{23} ,\;
b=c_{23}X+c_{12}c_{13}\alpha s_{12} -Ys_{23} \; .
\eea
The  eigenvalues of $H$ upto second order in $ \alpha $ are
\bea
 {{m^{'}_1}^2 \over 2 E}
 &\approx &\frac{\Delta m_{31}^2}{2E}\left[A+\alpha s_{12}^2+s_{13}^2  +
 \frac{|b|^2}{A}+\frac{|a|^2}{(-1+A)}\right],  \nonumber \\
{ {m^{'}_2}^2 \over 2 E}
&\approx &\frac{\Delta m_{31}^2}{2E}\left[\alpha c_{12}^2-\frac{|b|^2}{A}
\right], \;\; 
 {{m^{'}_3}^2 \over 2 E}
\approx \frac{\Delta m_{31}^2}{2E}\left[1- s_{13}^2+\frac{|a^\ast|^2}{(1-A)}\right]
\eea
In the same way we can calculate the eigenvalues keeping NSI in 23 block.
Using eqs. (4) and (13)  and putting $\e_{12}$ and $\e_{13}$
to zero one obtains the earlier known magic baseline condition \cite{magic1}
in presence of only SMI. 
For small $\sin \theta_{13} \leq \a $ this condition is given by
\be
 L={2n\pi}/({\sqrt{2}G_F n_e}).
\ee
The probability $P_{\nu_e \rightarrow \nu_\mu}$ of oscillation expression after using the baseline condition
in (4) for such small $\sin \theta_{13}$ is (upto order $\a^2$)
\bea
 P(\nu_e\rightarrow\nu_{\mu})\approx 4\frac{s^2_{23}}{(1-A)^2}|a|^2\sin^2\frac{\Delta m_{31}^2(1-A)L}{4E}
\eea
One can see from this probability expression that ${|a|}^2$ does not contain
CP violating phase $\delta$ when we consider only SMI but it does contain
$\delta$ when we keep NSI terms $X$ and $Y$ in $a$.
This means that the magic baseline condition (14) is not valid when there is
$\e_{12}$ and $\e_{13}$ as NSI. For brevity, we are
not showing the detailed calculation of obtaining the probability
expression if $\e_{22}$, $\e_{23}$ and $\e_{33}$ is considered. However,
one may
note that in such cases  probability $P(\nu_e \rightarrow \nu_{\mu})$
upto order $\alpha^2$  is same with that
in presence of only SMI. Any correction due to NSI is
present in  higher order of $\a$ only.
So for small $\sin \theta_{13}$, the magic baseline condition in (14)
is valid  when  NSI is present only in 23 block of $M$ in (9).

However, the above conclusions related to magic baseline condition
change if we consider large $\sin \theta_{13} > \a$.
Let us discuss the perturbative
approach for large $\theta_{13}$. This was considered in \cite{largetheta}
earlier for SMI only.  Here we use it for both SMI and NSI
particularly in the context of magic conditions.
We consider NSI in  12 and 13 elements.
Then $M$
in (9) can be written as $M= M^{(0)} + M^{(1)} + M^{(2)}$
where
\bea
 M^{(0)}&=&\frac{\Delta m_{31}^2}{2E}\pmatrix{A+s_{13}^2 & 0 &
 e^{-i\delta}s_{13}c_{13}\cr 0 & 0 & 0\cr e^{i\delta}s_{13}c_{13}
 & 0 & c_{13}^2},
\nonumber \\  M^{(1)}&=&\frac{\Delta m_{31}^2}{2E}\pmatrix{\alpha s_{12}^2
& b & a_{1}\cr b^\ast & \alpha c_{12}^2 & -e^{-i\delta}\alpha
c_{12}s_{12}s_{13}\cr a_{1}^{\ast} & -e^{i\delta}\alpha c_{12}s_{12}s_{13}
& 0},
\nonumber \\
 M^{(2)}&=&\frac{\Delta m_{31}^2}{2E}\pmatrix{-\alpha s_{13}^2s_{12}^2 & 0 & 0 \cr 0 & 0 & 0 \cr 0 & 0 & \alpha s_{13}^2s_{12}^2}
\eea
where
\bea
  a_{1}= c_{23}Y-e^{-i\delta}c_{13}s_{13}\alpha s_{12}^2 +Xs_{23}. 
\eea

The  eigenvalues upto second order correction in $\alpha$ are
\bea
 {{m^{'}_1}^2 \over 2 E}&\approx &\frac{\Delta m_{31}^2}{2E}
 \left[\frac{1}{2}(A+1-x)+\alpha s_{12}^2-\alpha s_{13}^2 s_{12}^2
 +\frac{2|b|^2}{A+1-x}-\frac{|a|^2}{x}\right], \nonumber
\\
{{m^{'}_2}^2 \over 2 E} &\approx &\frac{\Delta m_{31}^2}{2E}\left[\alpha c_{12}^2-\frac{2|b|^2}{A+1-x}-\frac{2(\alpha c_{12}s_{12}s_{13})^2}{A+1+x}\right]
, \nonumber \\ {{m^{'}_3}^2 \over 2 E}&\approx &\frac{\Delta m_{31}^2}{2E}\left[\frac{1}{2}(A+1+x)+\alpha s_{13}^2s_{12}^2+\frac{|a^\ast|^2}{x}+\frac{2(\alpha c_{12}s_{12}s_{13})^2}{A+1+x}\right]
\eea
where
\be
 x=(1+A^2-2A\cos2\theta_{13})^{1/2}.
\ee
In the same way we can calculate the eigenvalues keeping NSI terms in 23
block also. Using (18) the condition (4) may be written for large
$\sin \theta_{13} > \a $ as
\be
L =  8 E n \pi /\left\{ \Delta m_{31}^2 \left( A + 1 - x\right)\right\}
\ee
Unlike baseline condition in (14) the condition in (20) depends on
 $\theta_{13}$ (in $x$) and energy and also this does not give $\delta $ independent
 probability as discussed below.
For such large $\sin \theta_{13}$ the probability $P_{\nu_e \rightarrow \nu_\mu}$
with baseline condition in (20) is (upto order $\a^2$)
\bea
 P(\nu_e\rightarrow \nu_{\mu})\approx -4 Re[Z]\sin^{2}\frac{\Delta m^{2}_{31} L x}{8E}
-2Im[Z]\sin 2\frac{\Delta m^{2}_{31} L x}{8E}
\eea
where
\bea
 Z&=&\frac{s^2_{23}}{(1-\xi^2)}\left[-\xi^2 k^{2}-\frac{a_1 \xi^2
 k^{2}}{x}+\frac{a^\ast_1 \xi^3 k^{2}}{x}-\frac{|c_{23}Y+s_{23}X|^2
 \xi^4 k^{2}}{x^2}+\frac{|c_{23}Y+s_{23}X|^2 \xi^2 }{x^2}\right]
\nonumber
 \\
&+&\frac{c_{23}s_{23}}{(1-\xi^2)}\left[-\frac{4(c_{23}\b^{\ast}-
s_{23}\g^{\ast})\xi k}{(A+1+x)}\right]
\eea
and
\bea
 \b &=& C \; c_{23}+ B \; s_{23} \; ; \; \g = D \; c_{23} + C\; s_{23} \; ;
\; \xi=(-A+\cos2\theta_{13}+x)\mbox{cosec} 2\theta_{13}
 \; ; \nonumber \\
 \; k
 &= & 1/{\left[1+(-A+\cos 2 \theta_{13}+x)^2 \; {\mbox{cosec}}^{2} 2
 \theta_{13}\right]^{1/2}}
\eea
This probability
is not independent of $\delta$ due to the presence of $a_1$ in $Z$.
So it is not possible to get magic baseline condition (resulting in $\delta$
independent probability $P_{\nu_e \rightarrow \nu_\mu}$ upto order $\a^2$)
 for
large $\sin \theta_{13} > \a$ with or without NSI in any elements in $M$.

However,
if we consider some magic condition on neutrino energy then it is possible
to get $\delta$ independent probability $P_{\nu_e \rightarrow \nu_\mu}$ for both small and
large $\theta_{13}$
and also with and without NSI.
Using condition (6) and considering $P_{\nu_e \rightarrow \nu_\mu}$ upto
order $\a^2$ the magic energy condition for small $ \sin \theta_{13} \leq \a$
is written as
\be
 E=\Delta m_{31}^2/\left( \pm 4n\pi/L +2\sqrt{2}G_{F}n_{e}\right)\; .
\ee
Using the above energy condition for such small $\sin \theta_{13}$ with NSI terms
(upto order $\a^2$)
\be
 P(\nu_e \to \nu_\mu)
 \approx \frac{4c_{23}^2}{A^2}|c_{23}X+c_{12}c_{13}\alpha s_{12}-Ys_{23}|^2
 \sin^2\left( \frac{\Delta m_{31}^2AL}{4E}\right)\; .
\ee
With $X=Y=0$ this corresponds to Standard Model result. Unlike (15)
this is 
independent of $\delta$ even with NSI terms. This is one important
advantage of using magic energy condition instead of magic
baseline condition even for $\sin \theta_{13} \leq \a$ .
However, as the condition is on energy it might be useful
to consider monoenergetic neutrino beam as source \cite{mon, mono} to study such $\delta$ independent
probability. The other alternative way to study such probability
might be  to consider very small energy bins for neutrino energy which satisfies approximately
the above energy condition in (24).

Using condition (6) and considering $P_{\nu_e \rightarrow \nu_\mu}$ upto
order $\a^2$, the magic energy condition for large $ \sin \theta_{13} > \a $
is written as
\be
 E=\Delta m_{31}^2S\cos2\theta_{13}L^2\left({-1\pm \sqrt{1+Q/R}}\right)/\left(
 2 Q \right)
 \ee
where
\bea
 Q=(2n\pi)^2-S^2L^2 ;\;  
R=S^2 L^2 \cos^{2}2\theta_{13} ;\;
S=\sqrt{2}G_{F}n_e \nonumber .
\eea
Although this condition depends on $\theta_{13}$ but
with presently allowed values of $\theta_{13}$ \cite{tho}
this dependence is not so significant as shown later in Figure 1
in which $E$ satisfying condition (24) overlaps on $E$ satisfying
condition (26).
It is
important to note here  that unlike magic baseline
condition  (20,) this energy
condition (26) results in $\delta $ independent probability as shown below.
Using  condition (26) and considering NSI terms and $\sin \theta_{13}  > \a$,
the probability
$P_{\nu_e \rightarrow \nu_\mu}$
is  written as (upto order $\a^2$)
\bea
 P(\nu_e \to \nu_\mu)
\approx \frac{16|b|^2\xi^2c_{23}^2}{(1+\xi^2)(A+1-x)^2}
  \sin^2 \left( \frac{\Delta m_{31}^2L(A+1-x)}{8E}\right).
\eea
Unlike the energy condition (26), the probability of oscillation
depends significantly on $\theta_{13}$. 
This probability
is $\delta$ independent with or without NSI.

Due to   present ambiguity in the sign of
$\Delta m_{31}^2$, with (+) sign to $\Delta m_{31}^2$
 for  normal and
with (-) sign for inverted hierarchy of
neutrino masses, the energy conditions will be different. Apart from
hierarchy  sign
there is further consideration of choosing signs in energy conditions
as seen in (24) and (26). The requirement of
positive energy allows
considering both of those signs in the energy conditions
for normal hierarchy and
considering only (-) sign for inverted hierarchy.
For normal hierarchy for (-) sign in the conditions, $L > {\sqrt{2} n \pi
\over G_F n_e}$ but there is no bound for (+) sign.
For inverted hierarchy
$L < {\sqrt{2} n \pi \over G_F n_e}$.
Due to singularity at  $L = {\sqrt{2} n \pi \over G_F n_e}$,  neutrino
energy $E$ satisfying energy conditions are not possible at magic baseline
 length.

  Using magic energy condition
  one could  resolve
  ambiguities in $\delta - \theta_{13}$ (which usually happens for
  non-magic neutrino energy) as the probability is
$\delta$ independent. Furthermore, 
  using energy condition
one may also try
  to find the neutrino mass hierarchy. For illustration,
  let us consider say
  nature admits normal hierarchy and $\sin \theta_{13} \leq \a $ and
  for simplicity say NSI is absent.
  Certain neutrino energy has been
   fixed
  by energy condition in (24) with appropriate choice of $n$ value for
  which monoenergetic neutrino will be feasible in experiment.
Under such conditions 
  the probability
  in (25) is independent of $\theta_{13}$ upto order $\a^2$.
  So the probability
  has fixed value and normal hierarchy could be verified by experiment upto
  order $\a^2 $ from the number of $\mu $ events observed at the detector.
  This number in general, differs from that which one could have obtained
  for inverted hierarchy in this case. The reason is that, for
  inverted hierarchy the same neutrino energy will not correspond to
  magic energy anymore. In  fact, in this case, one can show that 
the difference in probability of oscillation
$P_{\nu_e \rightarrow \nu_{\mu}}^{N(magic)}$ for normal hierarchy with magic energy condition
 with  that for inverted hierarchy
 $P_{\nu_e \rightarrow \nu_\mu}^{I(non-magic)}$
without
 any magic energy
 condition is
 \bea
 P_{\nu_e \rightarrow \nu_\mu}^{N(magic)}-
 P_{\nu_e \rightarrow \nu_\mu}^{I(non-magic)} &=& - {1 \over (1 + A)^2 }
 \sin^2 2 \theta_{13} \sin^2 \theta_{23}
 \sin^2 \left( { (1+ A)
 \Delta m_{31}^2 L \over {4 E} } \right) \nonumber \\ &&+
 {1 \over {A (1+A)}} \a
 \sin 2 \theta_{13} \cos \theta_{13} \sin 2 \theta_{12} \sin 2 \theta_{23}
 \times \nonumber \\
 &&\cos \left(\delta-
 {\Delta m_{31}^2 L \over {4 E}} \right)
 \sin
 \left({A \Delta m_{31}^2 L \over {4 E}}\right) \sin
 \left( {(1+A) \Delta m_{31}^2 L \over {4 E}}  \right)
 \nonumber
 \eea
which does not vanish in general, for any value of $\delta $ unless
 $\sin \theta_{13}$ vanishes. However, after fixing $L$ and $E$ for
 the experiment one may check whether this difference vanishes or not.
 So for normal hierarchy,
in general, one is supposed to get different number of $\mu $
events at the detector than that for inverted
hierarchy and neutrino mass hierarchy may be resolved for $0 < \sin \theta_{13} \leq  \a
$. If the number of $\mu $ events does not match with the expected one
for normal hierarchy at magic energy, one may try the magic energy
for inverted hierarchy in the experiment. Similar to the above expression,
one can show that
$ P_{\nu_e \rightarrow \nu_\mu}^{I(magic)}-
 P_{\nu_e \rightarrow \nu_\mu}^{N(non-magic)}$  will not vanish in general, for any value
 of $\delta $ unless $\sin \theta_{13}$ vanishes. 

 For larger $\sin \theta_{13} > \a $, the above difference has more
complicated form. Same method can be adopted in this case also
 to resolve hierarchy, provided that the difference
of $ P_{\nu_e \rightarrow \nu_\mu}^N$ (for normal hierarchy) and
 $P_{\nu_e \rightarrow \nu_\mu}^I $ (for inverted hierarchy)
  does not vanish for $L$ and $E$ value considered
 in the experiment (where $E$ could be magic energy for either normal
 or inverted hierarchy). In case, it vanishes either for particular
 combination of $\sin \theta_{13}$ or $\delta $ one may consider
 for the same baseline, a different neutrino magic energy by changing
 $n$ value in the magic energy
 condition for which such difference may not vanish.

We now, illustrate the use of such magic energy conditions to get $\delta $ independent
probability and discuss the experimental feasibility. One is required
to fine-tune the energy of the monoenergetic
neutrino beam. In the electron capture facility as discussed at the
beginning, the neutrino energy may be fixed by appropriately choosing the
boost of the ion source.
We consider the monoenergetic neutrino
beam for the ion type $^{150}Dy$  with neutrino energy $E_r$
at rest given by 1.4 MeV
as suggested in ref. \cite{mon}. We have chosen Lorentz boost $\gamma$ in the range of
90 -195 such that the magic energy condition is satisfied. The neutrino
energy $E$ is fixed in the forward direction by the boost as
$E = E_r \gamma $. We have assumed
flux of $10^{18}$
neutrinos per year.
We are considering a baseline
of length 650 Km from CERN to  megaton water Cerenkov detector possibly
located at
Canfranc in Spain. For such baseline the constant matter density
has been approximated to be $ 4.21 $ gm/cc. 
For our subsequent sensitivity analysis of oscillation parameters
we mention here the detector characteristics also for such experimental set-up.
\cite{magic3}: (a) Fiducial mass = 500 Kton  (b) Detection efficiency
($\epsilon $ ) = 50 \% (c) Charge identification  efficiency ($I_e$) = 95 \%.

\begin{figure}[htb]
\begin{center}
\includegraphics[width=6cm]{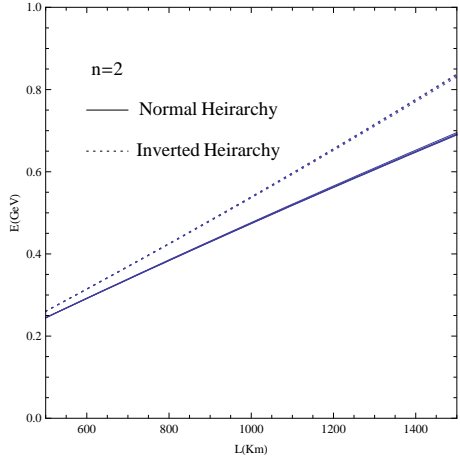}

\vspace{0.1cm}

\caption[] {{\small{ Plots for neutrino energy ($E$) versus length
($L$) of
baseline
using energy conditions (24) for any $sin \theta_{13} \leq \a  $
and  (26) for $\theta_{13}=5, 8, 12  $ for $n=2$ for both hierarchies.
 }}}
\end{center}
\end{figure}

In Figure-1 we have shown the energy versus length of baseline satisfying
magic energy conditions. Condition (24) has been considered
 for any $\sin \theta_{13}  \leq \a $
and condition
in (26) for
$\theta_{13} =  5, \; 8, \; $ and 12 degrees.
However, it is seen from the figure that the plots with different
$\theta_{13}$   are almost overlapping with each other indicating
very small change in energy $E$  with $L$ due to variations of unknown
parameter $\theta_{13}$.
In plotting Figure 1, instead of $n=1 $ we have considered
$n=2$ in the
energy conditions
(24) and (26) so that for the above-mentioned baseline of length 650 Km, the
magic neutrino energy lies in experimentally feasible  range.

\begin{figure}[htb]
\begin{center}
\includegraphics[width=6cm]{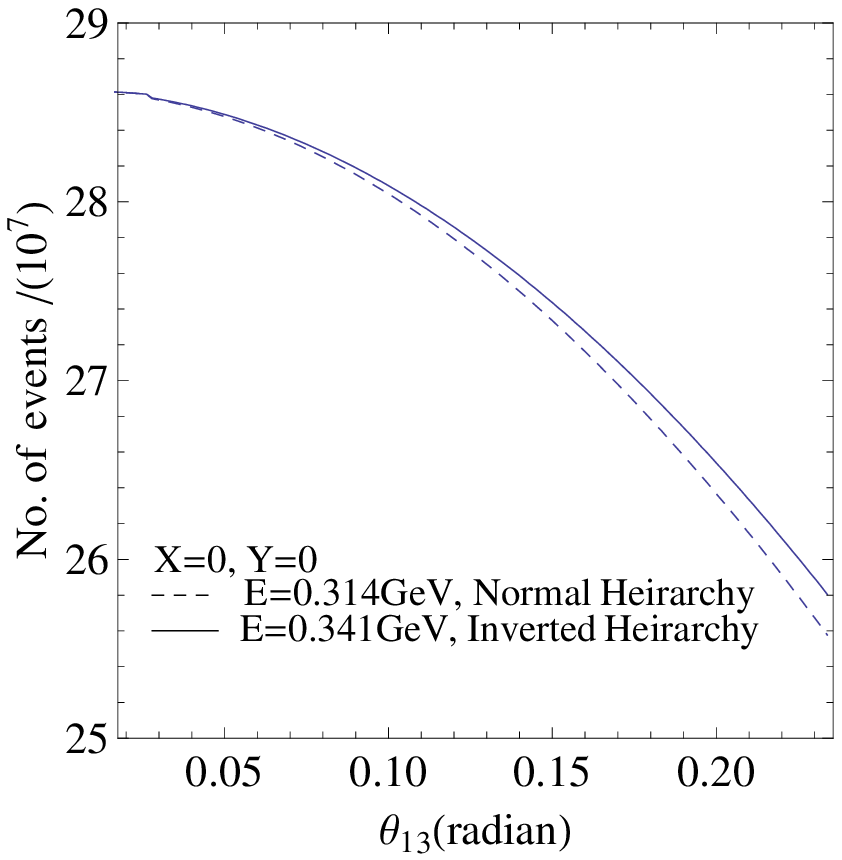}

\vspace*{-3mm}
\caption[] {{\small Number of $\mu $ events
versus mixing angle $\theta_{13}$ with 5 years running without
NSI (i.e $X=Y=0$).
}}
\end{center}
\end{figure}

\begin{figure}[htb]
\begin{center}
\includegraphics[width=5cm]{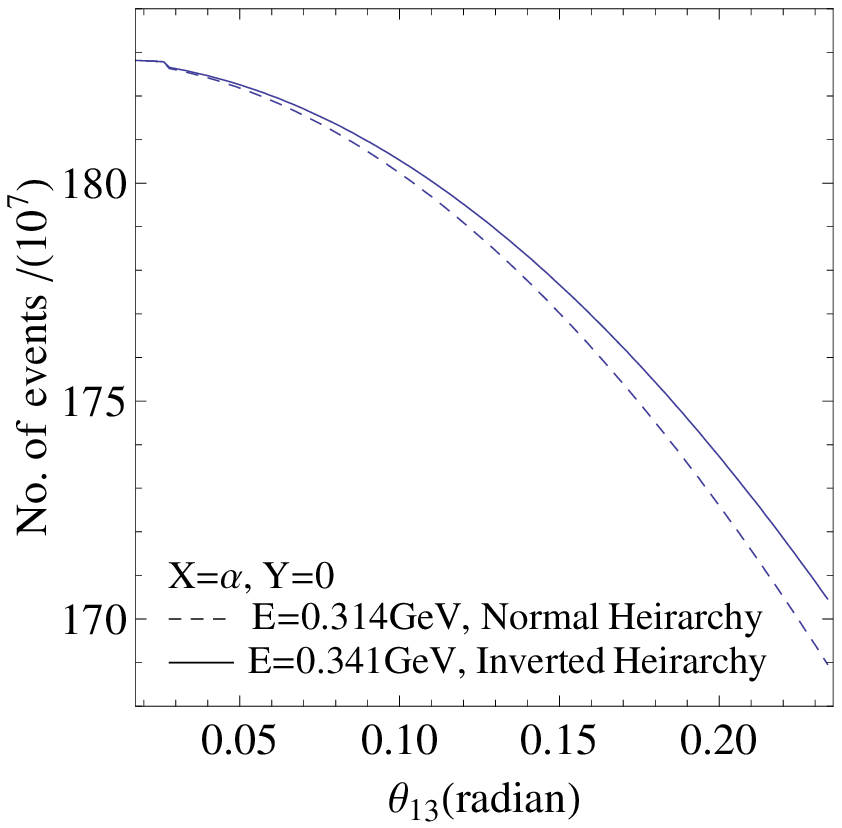}
\hspace{3mm} 
\includegraphics[width=5cm]{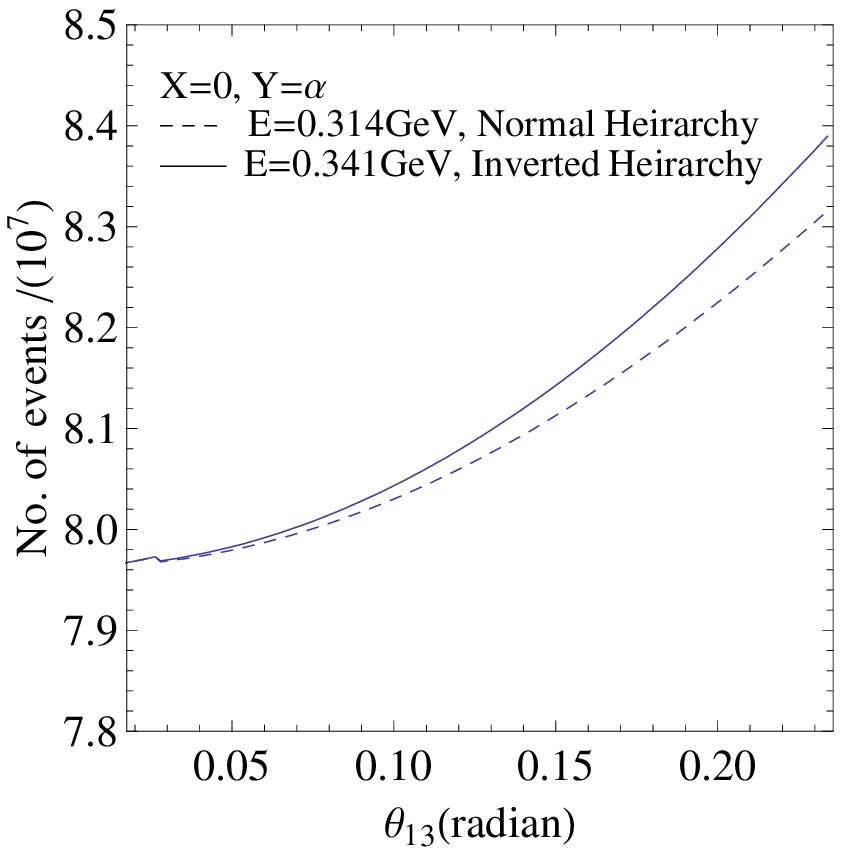}

\hspace{1.7cm}(a)\hspace{5.6cm} (b)
\vspace*{-2mm}
\vspace*{-3mm}

\vspace*{1mm}
\caption[] {{\small Number of $\mu $ events
versus mixing angle $\theta_{13}$ with 5 years running with NSI  $X$ and  $Y$.
}}
\end{center}
\end{figure}

\begin{figure}[htb]
\begin{center}
\includegraphics[width=5cm]{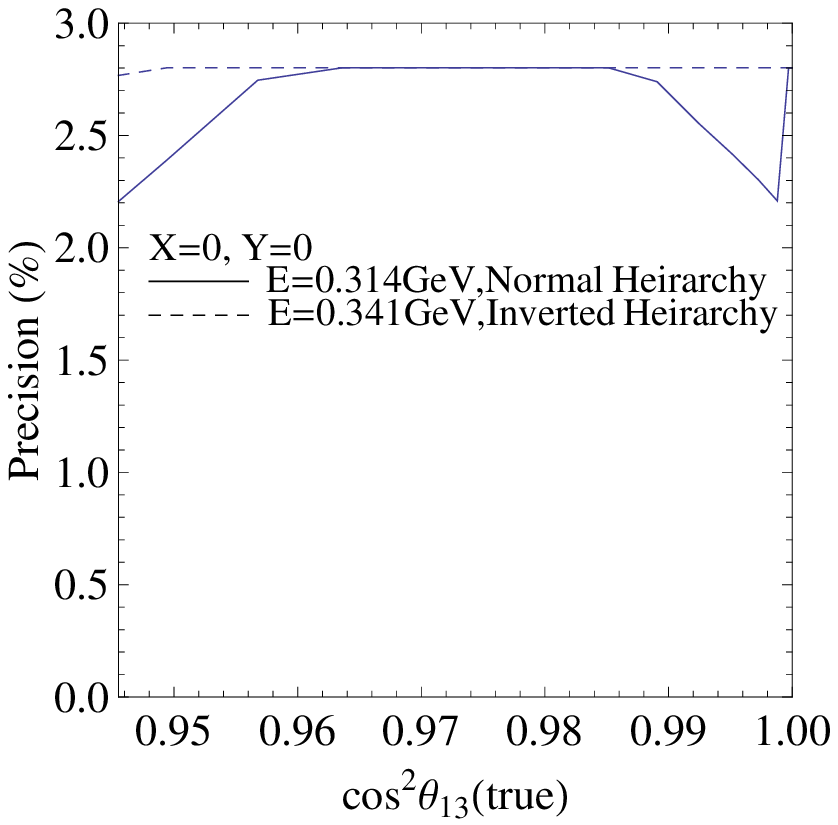}
\hspace{2cm}
\includegraphics[width=5cm]{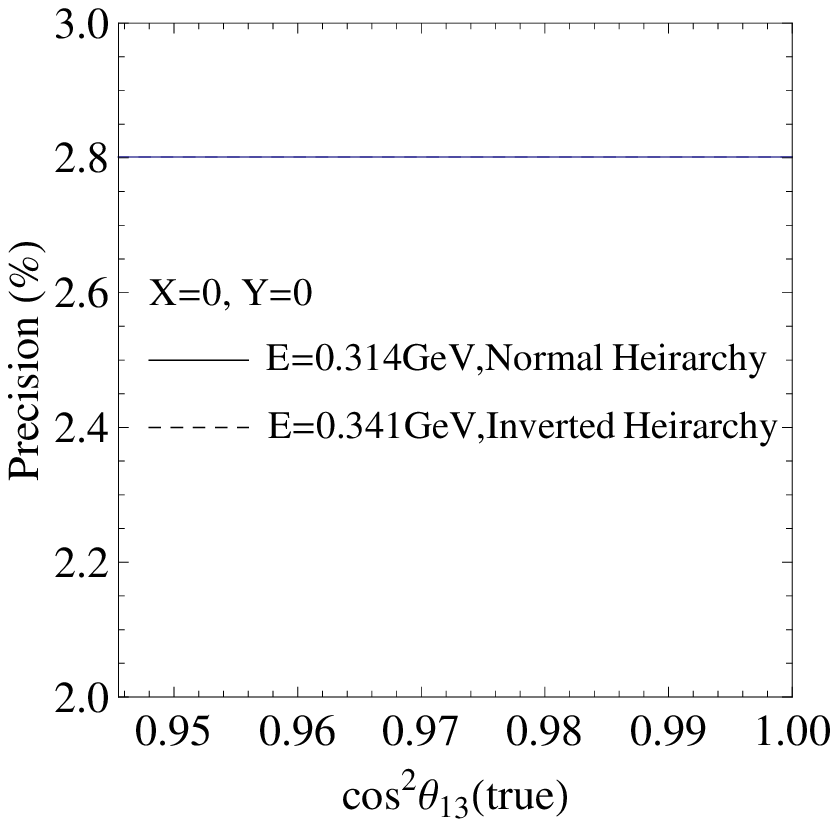}


\hspace{1.1cm}(a)\hspace{6.5cm} (b)
\vspace*{-2mm}
\vspace*{-3mm}
\caption[] {{\small Precision in the measurement of $\cos^2  \theta_{13} $
 without NSI expected  with 5 years running at $1 \sigma$ in (a) and
at $3 \sigma $ in (b).  }}
\end{center}
\end{figure}

\begin{figure}[htb]
\begin{center}
\includegraphics[width=5cm]{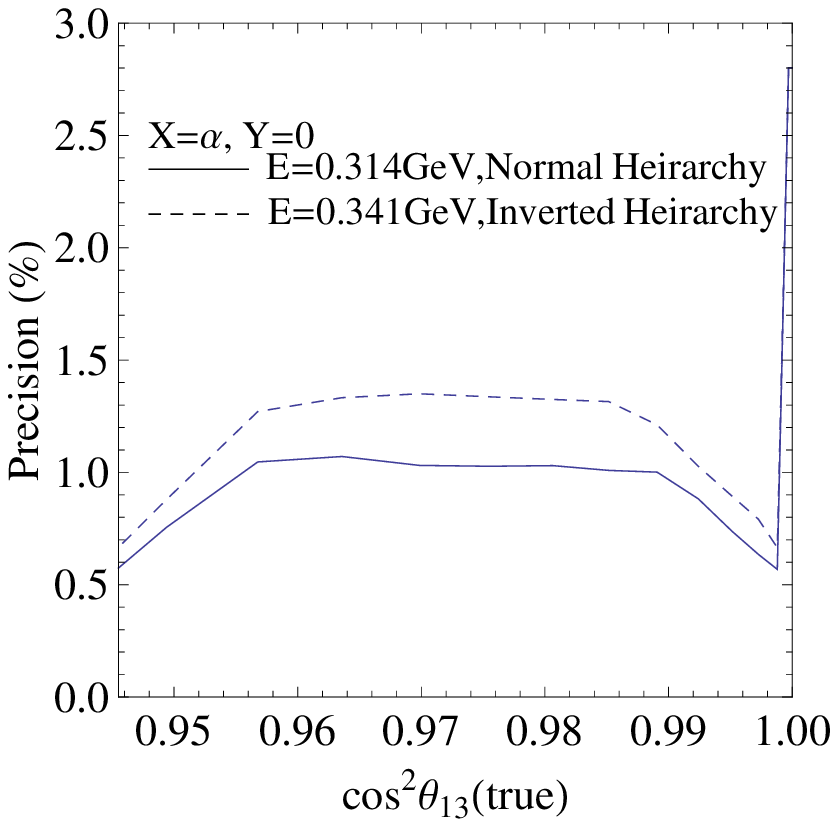}
\hspace{2cm}
\includegraphics[width=5cm]{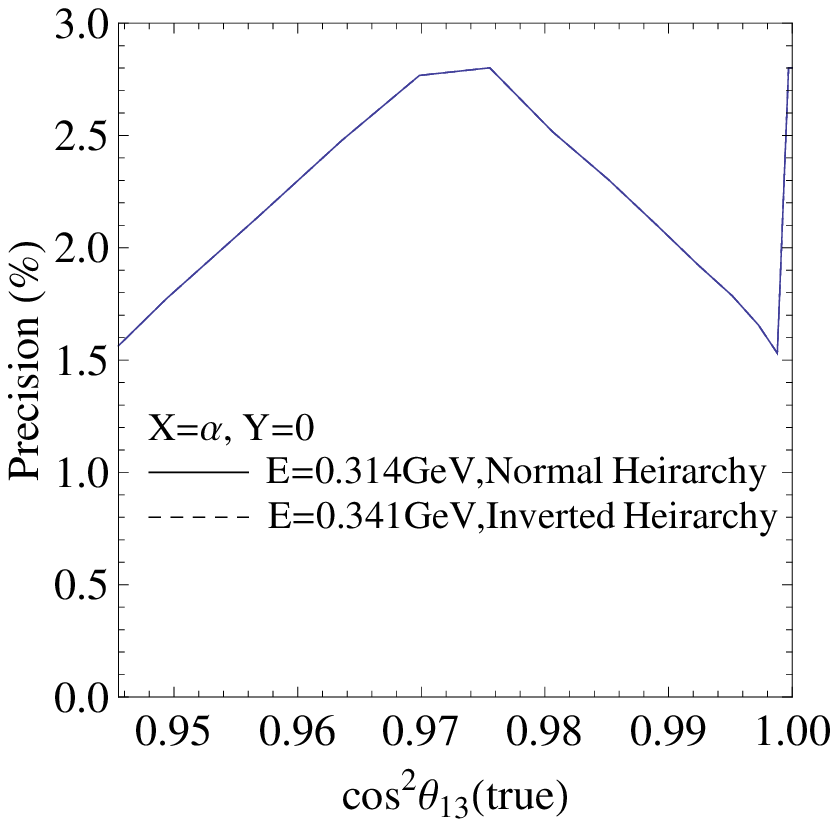}


\hspace{1.1cm}(a)\hspace{6.5cm} (b)
\vspace*{-2mm}
\vspace*{3mm}

\includegraphics[width=5cm]{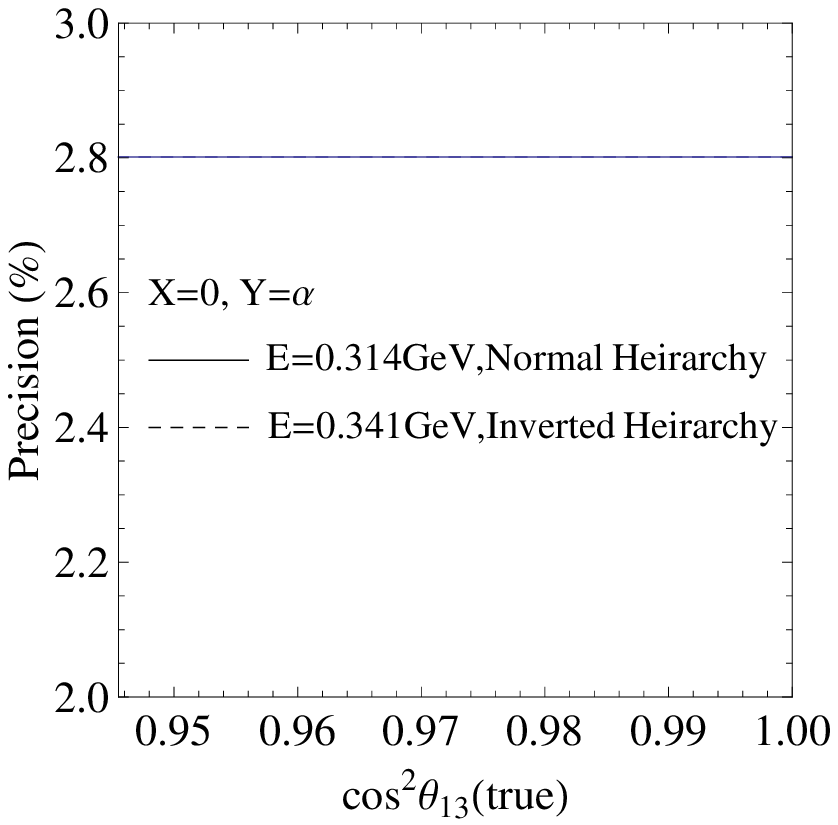}
\hspace{2cm}
\includegraphics[width=5cm]{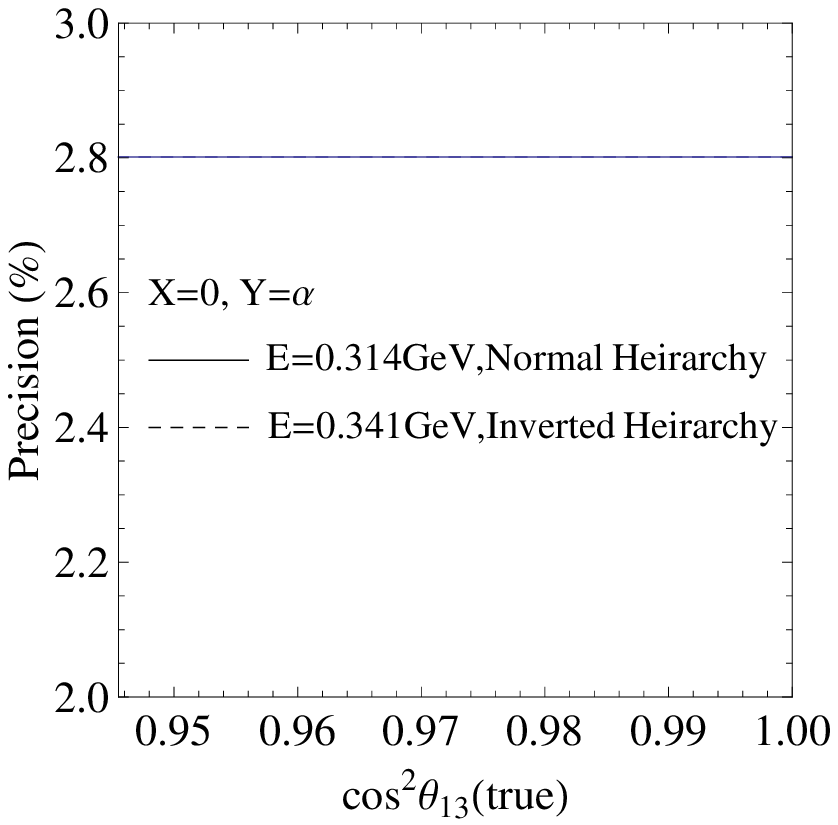}


\hspace{1.1cm}(c)\hspace{6.5cm} (d)
\vspace*{-2mm}
\vspace*{-3mm}
\caption[] {{\small Precision in the measurement of $\cos^2  \theta_{13} $
 with NSI expected with 5 years running at $1 \sigma$ in (a) and
(c) and at $3 \sigma $ in (b) and (d).  }}
\end{center}
\end{figure}

Finally we discuss the sensitivity in measuring the unknown oscillation parameters
like $\theta_{13}$ and $\Delta m_{31}^2$ in the experimental set-up with
monoenergetic neutrino beam \cite{mon, mono}. The number of $\mu $ events
expected at the detector due to $\nu_e \rightarrow \nu_\mu $ oscillation is
given by 
\bea
N_{\mu} = T \; n_n \; I_e \; \e \; E \; \phi (E) \; \
\sigma_{\nu_\mu} (E)
P_{\nu_e \rightarrow \nu_\mu} (E) 
\eea
where $T=$  time period , $n_n=$  number of target nucleons, $\phi (E) =$
flux, $\sigma_{\nu_\mu} (E)=$ detection cross-section. 
As we are considering monoenergetic neutrino beam so the
energy resolution function on which normally number of events depends,
may be considered to be effectively 1 and as such is not mentioned in eq. (28).
In figure 2 we have shown for both hierarchies the variation of the
number of events expected for a priod of 5 years with
$\theta_{13}$ in absence of any NSI  and with NSI respectively. The number of
 $\mu $ events are quite large as neutrinos
have some fixed energy instead of   Gaussian distribution  of energy.
Number of events with
inverted hierarchy is found to be slightly higher  than that for
normal hierarchy for different values of $\theta_{13}$ with or without
NSI.

In plotting figures 4  and 5 we define $\chi^2_{total}$ as
\bea
\chi^2_{total} = {\left( {N^{expt} - N^{th} \over \Delta N 
 } \right) }^2
+ {\left( { |\Delta m_{31}^2 | - |\Delta m_{31}^2(true) | \over
\sigma
\left( \Delta m_{31}^2 \right) } \right)  }^2 +
{\left( {\sin^2 2 \theta_{23} - \sin^2 2 \theta_{23}(true) \over
\sigma \left( \sin^2 2 \theta_{23}  \right)  }  \right) }^2
\eea
where  $N^{expt}$ and $N^{th}$ stands for experimental and theoretical
number of  $\mu$ events
respectively at the detectors and the error
in $N^{th}$ = $\Delta N
= \Delta N_{pert} + \Delta N_{\a^2}$. Here, $\Delta N_{pert} $ and
 $\Delta N_{\a^2}$ are the differential
change in $N^{th}$ considering perturbative expression of $P_{\nu_e
\rightarrow \nu_\mu}$ in (27) and the perturbative error of order $\a^2$ in
(27) respectively. The $\Delta N_{\a^2}$ takes care of maximum possible
correlated error
due to any true value of $\delta $ and we have assumed democratic form of correlation matrix.
Unlike
magic baseline condition for magic energy condition the
perturbative expression
of probability depends on $\cos^2 \theta_{13} $ instead
of $\sin^2 2 \theta_{13} $ and thus
  resulting in large number of events ($  \sim 10^7 $) even for
  small $\theta_{13}$. So
unlike \cite{magic3} instead of Poisson distribution, we have
considered Gaussian distribution for error in  $N$.
In evaluating 2nd
and 3rd terms   on right side of  eq.(29) we have considered
$\Delta m_{31}^2(true) = 2.5 \times 10^{-3}\; eV^2$,
$\sin^2 2 \theta_{23}(true)=1.0$,
$\sigma
\left( \Delta m_{31}^2 \right)  =1.5 \% $,
$\sigma
\left( \sin ^2 2 \theta_{23} \right)  = 1 \% $. Based on
various experimental data set
\cite{magic3}
the following $3 \sigma $ constraints
on the following oscillation parameters have been considered :
\bea
&&2 \times 10^{-3} \; eV^2 < | \Delta m_{31}^2 |  < 3.2 \times 10^{-3} \; eV^2 ;
\nonumber \\
&&35.67^\circ  < \theta_{23} < 55.55^\circ .
\nonumber
\eea

In figure 4 and 5 we have shown the precision in measuring
$\cos^2 \theta_{13}$ which appears in the expression of
$P_{\nu_e \rightarrow \nu_\mu }$ satisfying magic energy condition.
This is defined by
\bea
 Precision = {
\cos^2 \theta_{13} (min)-
\cos^2 \theta_{13} (max)  \over 
{\cos^2 \theta_{13} (min)  +
\cos^2 \theta_{13} (max)}} 100\%
\eea
in which $\cos^2 \theta_{13} (min) $ and
$\cos^2 \theta_{13} (max)$  are the smallest and largest values respectively
of $\cos^2 \theta_{13}$ at the given confidence level. We have shown the precision
at $1 \sigma  $ and $ 3 \sigma $ level with and without NSI
for both the hierarchies as mentioned in the figures.
In finding precision in measurement of $\cos^2 \theta_{13}$ in presence
of NSI, we have assumed that the strength of NSI couplings
are known from some other experiments. The number of muon events at the detector
changes when such NSI are included and thus changes the level
of precision in measurement of $\cos^2 \theta_{13}$. The precision as
defined above is better for its lower values. In  general in both
figures 4 \& 5 the precision initially deteriorates with increase
of $\cos^2 \theta_{13} $
values and after reaching some limiting values of $\cos^2 \theta_{13}$
the precision starts improving and finally for larger values of
$\cos^2 \theta_{13}$ again it deteriorates. However, in some cases the
change in precision is insignificant with the variations of
$\cos^2 \theta_{13}$ as seen in figures 4(b) and 5 (c) \& (d). From figures
the differences in precision
for normal and inverted hierarchies
are particularly found at $1 \sigma $ level except in figure 5 (c) where
non-zero value of $Y$ as NSI has been considered.
Comparing three cases
- (a) no NSI (b) NSI with only
non-zero $X$ (c) NSI with only non-zero $Y$,
it is seen from figures 4 \& 5 that
the precision in measurement of $\cos^2 \theta_{13}$ is best for (b)
and the precision is better for (a) than that for (c). Actually more the number
of $\mu$ events the better is the precision as these cases may be seen
from figures 2 \& 3.

As concluding remarks we  mention that to get
$P_{\nu_e \rightarrow \nu_\mu}$
almost independent of
unknown $CP$ violating phase $\delta $ one may
consider either  magic baseline condition
on the length of baseline or  the magic neutrino energy condition.
However, there are some disadvntages in considering the
magic baseline condition which are not present when magic energy condition
is considered. The magic baseline condition exists only
for small $\sin\theta_{13} \leq \a$. Also this condition
exists when NSI are considered in only 23
block of effective neutrino mass matrix $M$
(as
$P_{\nu_e \to \nu_\mu}$ upto order $\a^2$ is independent of NSI in 23 block).
However, magic baseline condition is not possible if NSI terms
are present in 12 , 13 elements of $M$.
For large $\sin\theta_{13} > \a $ using magic baseline
condition in (20)
 it is not possible to get $\delta $ independent probability
$ P_{\nu_e \to \nu_\mu}$ 
 with or without NSI. Magic baseline condition
 will also depend on neutrino energy
 for
$\sin \theta_{13} > \a$. 
 Also to
 place neutrino detector at a location satisfying magic baseline condition
 may not be always feasible.
Some of the drawbacks mentioned above in considering magic baseline condition
may be overcome by considering the  condition on
neutrino energy. Under magic neutrino energy condition for both small and
large $\sin \theta_{13}$ and also
with or without NSI one gets $\delta$
independent
probability $P_{\nu_e \rightarrow \nu_\mu}$.

Using energy condition there is  scope to find out the hierarchy of neutrino masses
and to obtain overall good precision in the measurement of
$\cos^2 \theta_{13}$ over the allowed range of $\theta_{13}$ as discussed
earlier.
Depending on the presence or absence of NSI, the number of $\mu $ events could
 significantly
differ as shown in Fig. 2 \& 3 and could signal the presence of new physics.
If the experimental data indicates the presence of NSI then to find
NSI as well as $\theta_{13}$ one may consider changing the 
magic neutrino energy by changing the Lorentz boost in the same experimental set-up.
Then NSI terms like
$\e_{12}$ and $\e_{13}$
as well as $\theta_{13}$  may be known from $ P_{\nu_e \to \nu_\mu} $ in (25) for $\sin
\theta_{13} < \a$ or (27) for $\sin \theta_{13} > \a$ after matching those
probabilities
with experimntal data on number of $\mu $ events using eq. (28).
In long baseline experiments, 
 monoenergetic neutrino beam as source with neutrino energy satisfying
 magic energy condition,  could be
highly useful in future advanced precision measuments of neutrino oscillation
parameters, in resolving hierarchies of neutrino masses and  also in
 searching NSI of neutrinos with matter.

{\bf Acknowledgment:} Both Z. R. and R. A. like to thank Amitava Raychaudhuri for
discussion and also like to acknowledge the hospitality provided
by Harish-chandra Research Institute, Allahabad, India under DAE XI-th
plan project on `Neutrino Physics'.

\end{document}